\documentstyle[epsf,psfig,aps]{revtex}
\begin{document}
\tighten
\title{Coherent medium as a polarization splitter of pulses}
\author{G. S. Agarwal and Shubhrangshu Dasgupta}
\address{Physical Research Laboratory, Navrangpura, Ahmedabad - 380 009, India}
\date{\today}

\maketitle

\begin{abstract}

We show how one can use the anisotropic properties of a coherent medium to 
separate temporally the two polarization components of a linearly polarized 
pulse. This is achieved by applying a control field such that one component 
of the pulse  becomes ultraslow while the other component's group velocity 
is almost unaffected by the medium. We present analytical and numerical 
results to support the functioning of such a coherent medium as a 
polarization splitter of pulses.

\end {abstract}

One of the remarkable consequences of one's ability to manage the dispersion
 of an optical medium \cite{SPT86,harris1} has been the manipulation of the optical pulses. While 
the dispersion management has been extensively practised in nonresonant 
systems like fibers, its use in the context of propagation in resonant 
systems is gaining prominence only in recent times. In particular one has 
discovered ultraslow light \cite{slow1,Hau}, superluminal propagation 
\cite{super1,super2}, stoppage \cite{stopOlga} and storage \cite{storePhillips} 
of light. The atomic coherences and quantum interferences \cite{harris2} are the key 
elements in control of optical properties of a medium. 

Here we examine the propagation of a linearly polarized pulse in an anisotropic medium. 
It is well known that an applied magnetic field makes a medium anisotropic 
leading to the Faraday effect \cite{faraday}. Thus
the polarization of a cw field is rotated as if the field propagates through 
an isotropic medium with a magnetic field applied in the direction in which the cw field is propagating. Some earlier 
theoretical \cite{mor,conn} and experimental \cite{gaeta} studies have shown 
how this magneto-optical 
rotation of a cw field in a medium can be enhanced to a large extent by using control 
lasers. One even found newer regions of frequency where the enhancement of 
the rotation angle was large \cite{mor1}. Further applications of coherent control of an anisotropic medium have been suggested \cite{fresnel}. 

In this paper we discuss an important new 
application of coherent control in an anisotropic medium. We show how an anisotropic 
coherent medium can be used to separate out two polarization components of a pulse.
The idea was to use a control laser 
appropriately polarized and of suitable frequency so that one of the two 
circularly polarized components of the linearly polarized pulse propagates almost without absorption and its dispersion becomes quite different. The other component, being detuned, has only small absorption. Thus if we were to think of the pulse as a combination 
of two polarized components, then one component propagates as if the medium 
were transparent and the dispersion for this component is such that it 
becomes ultraslow. The other component propagates without much effect as 
the medium is nonresonant. Clearly under these conditions the two 
polarization components of the pulse separate out in time. Thus the medium 
would act like a polarization splitter of the pulses.

Consider the propagation of a linearly polarized 
laser pulse through an anisotropic medium of length $L$. Let us write the input pulse
in terms of its Fourier components as
\begin{equation}
\vec{E}(z,t)=\hat{x}\int_{-\infty}^{\infty}{\cal E}(\omega)\exp{\left\{\{i\omega\left(\frac{z}{c}-t\right)\right\}}d\omega + \textrm{c.c.},
\label{input}
\end{equation}
where we assume that the pulse has a small spectral width. 
The amplitude $\hat{x}{\cal E}$ can be resolved in terms of two circular components
\begin{equation}
\hat{x}{\cal E}=\hat{\epsilon}_{+}{\cal E}_{+}+\hat{\epsilon}_{-}{\cal E}_{-},~~~{\cal E}_\pm={\cal E}/\sqrt{2},
\label{circ}
\end{equation}
where, unit orthogonal polarization vectors $\hat{\epsilon}_{\pm}$ correspond to $\sigma^{\pm}$ polarizations, and are given by 
\begin{equation}
\hat{\epsilon}_{\pm}=\frac{1}{\sqrt{2}}(\hat{x}\pm i\hat{y}).
\label{comp}
\end{equation}
The induced polarization in the medium due to the interaction with 
the linearly polarized probe can be expressed as
\begin{equation}
\vec{P}(z,t)=\hat{\epsilon}_+P_+(z,t)+\hat{\epsilon}_-P_-(z,t),
\label{pola}
\end{equation}
\begin{equation}
P_\pm(z,t)=\int_{-\infty}^{\infty}\chi_\pm(\omega){\cal E}_\pm(z,\omega)e^{-i\omega t}d\omega.
\label{polc}
\end{equation}
Here, $\chi_{\pm}(\omega)$ are the complex susceptibilities for the two circularly polarized components inside the medium. 

If we assume that the density of the medium is small so that the back reflections 
are negligible, then the field at the output can be written as 
\begin{eqnarray}
\vec{E}(L,t)&\equiv & \hat{\epsilon}_+\int^{+\infty}_{-\infty}d\omega\;{\cal E}_+(\omega)\exp{\left\{i\omega \left(\frac{L}{c}-t\right)+\frac{2\pi i\omega L}{c}\chi_+(\omega)\right\}}\nonumber\\
&+&\hat{\epsilon}_-\int^{+\infty}_{-\infty}d\omega\;{\cal E}_-(\omega)\exp{\left\{i\omega \left(\frac{L}{c}-t\right)+\frac{2\pi i\omega L}{c}\chi_-(\omega)\right\}}.
\label{out}
\end{eqnarray}

Clearly the two components of the pulse will travel with different group 
velocities $v_g^\pm$ given by 
\begin{equation}
v_g^\pm = c/n_g^\pm ; ~~~~~n_g^\pm \approx 1+2\pi \chi_\pm (\omega) +2\pi \omega \frac{\partial \chi_\pm (\omega)}{\partial \omega},
\label{ng}
\end{equation}
where the expression (\ref{ng}) is to be evaluated at the central frequency of the pulse.

\begin{figure}
\epsfxsize 9cm
\centerline{
\epsfbox{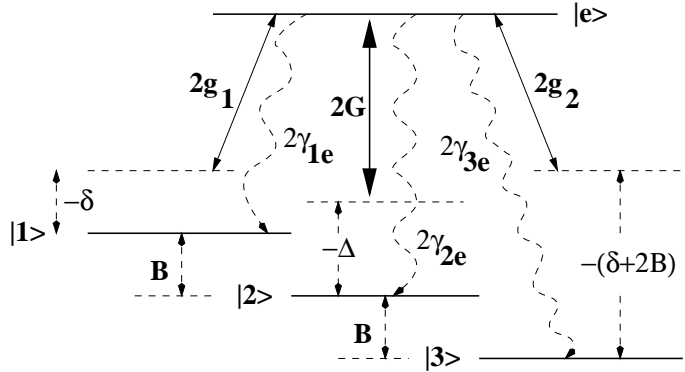}}
\caption{The atomic level configuration for splitting an input pulse temporally.
 The level $|e\rangle$ is coupled to $|1\rangle$ and $|3\rangle$ by $\sigma^-$ 
and $\sigma^+$ polarized components of the input pulse, 
with respective Rabi frequencies $2g_1$ and $2g_2$ and both with central frequency 
$\omega_0$. A pump field with Rabi 
frequency $2G$ couples the level $|e\rangle$ to $|2\rangle$. Here $B$ is the
Zeeman separation between the degenerate states. The $\sigma^-$ component 
is detuned from the corresponding transition frequency by $\delta$ and
the pump detuning is $\Delta$. $2\gamma_{je}$ $(j=1,2,3)$ are the spontaneous 
decay rates from $|e\rangle$ to $|j\rangle$.}
\label{config}
\end{figure}

We will now demonstrate how the ideas of coherent control can be used to separate temporally the two components of the pulse. The idea is to produce large anisotropy
between $n_g^+$ and $n_g^-$.
We consider a generic four-level model as shown in Fig.\ \ref{config} for this purpose. The relevant 
energy-levels are found in many systems such as in $^{23}$Na \cite{renzoni}, $^7$Li \cite{zeil}, and Pr:YSO \cite{ham}. The upper level which is a $|M_F=0\rangle$ state is coupled to the ground levels by a laser probe. The degeneracy of the
 ground level has been removed by applying a static magnetic field of strength $B$ in the direction of the propagation of the applied laser fields, as in the case of Faraday effect \cite{faraday}. The orthogonal components of the probe 
with $\sigma^-$ and $\sigma^+$ polarizations interact with $|e\rangle \leftrightarrow  |1\rangle (M_F=+1)$ and $|e\rangle \leftrightarrow |3\rangle (M_F=-1)$ 
transitions, respectively. Renzoni {\it et al.\/} \cite{arimondo} used the same atomic configuration to investigate the possibility of coherent population trapping using cw field of arbitrary intensities. In a dressed state approach they have shown that the long-interaction-time-evolution of the system can be completely characterized by the effective line-width of the noncoupled state. 

For the present configuration, the two circular components have the following susceptibilities assuming that the applied pulse is weak so that the medium behaves like a linear medium:
\begin{equation}
\chi_+(\omega)=\left(\frac{ND^2}{\hbar\Gamma}\right)\frac{-i\Gamma}{2[i(\delta+2B)-\Gamma_{e3}]};\;\;\;\;\;\chi_-(\omega)=\left(\frac{ND^2}{\hbar\Gamma}\right)\frac{-i\Gamma}{2(i\delta-\Gamma_{e1})},
\label{noG}
\end{equation}  
where $N$ is the atomic number density of the medium, $D$ is the magnitude of the dipole
moment matrix element between the levels $|e\rangle$ and $|1\rangle$, the pulse detuning $\delta$ is defined as $\delta=\omega-\omega_{e1}$,  
 $\omega_{ej}$ $(j=1,2,3)$ is the atomic 
transition frequency between the levels $|e\rangle$ and $|j\rangle$,
$\Gamma_{ej}=\Gamma=6\gamma$ is the 
decay rate of the off-diagonal element of the density matrix  between the levels $|e\rangle$ and $|j\rangle$, and $\gamma=A/12$, $A$ being the total spontaneous emission rate of the level $|e\rangle$.
The Rabi frequencies for the corresponding transitions are defined as 
\begin{equation}
2g_1=2\frac{\vec{d}_{e1}.\hat{x}{\cal E}}{\hbar}=\frac{D{\cal E}_-}{\hbar};~~~2g_2=2\frac{\vec{d}_{e3}.\hat{x}{\cal E}}{\hbar}=-\frac{D{\cal E}_+}{\hbar},
\label{rabi}
\end{equation}
where, $D$ is proportional to the reduced matrix elements for the relevant $|F_e,M_F=0\rangle \leftrightarrow |F_g,M_F=\pm 1\rangle$ transitions
and  can be calculated using the Wigner-Eckart theorem for the hyperfine levels \cite{sobel}. 
Note that, the magnetic field applied makes the system anisotropic, as $\chi_\pm$
are different [Eq.\ (\ref{noG})]. Thus, the medium will separate the input pulse into two 
orthogonal components provided we work in a region of frequencies where absorption is small.
Further in order to produce considerable pulse separation we have to work in a region 
so that there is large asymmetry between $\chi_+$ and $\chi_-$. This requires very large
magnetic fields, which could create a Paschen-Back splitting in both the excited and ground states \cite{xu}. In order to overcome these difficulties and to produce
very significant temporal separation between the two circularly polarized components of the pulse, we use the electromagnetically induced transparency (EIT). We apply a 
coherent cw field on the transition $|e \rangle \leftrightarrow |2\rangle $ 
\begin{equation}
\vec{E}_c(z,t)=\vec{\cal E}_c(z)e^{-i\omega_c t}+\textrm{c.c.}
\end{equation}
The application of this coherent field modifies the susceptibilities to $\bar{\chi}_\pm(\omega)$:
\begin{mathletters}
\begin{eqnarray}
\bar{\chi}_+(\omega)&=&\left(\frac{ND^2}{\hbar\Gamma}\right)\frac{1}{2}\frac{-i\Gamma[i(\delta+2B-\Delta)-\Gamma_{23}]}{[i(\delta+2B)-\Gamma_{e3}][i(\delta+2B-\Delta)-\Gamma_{23}]+|G|^2},\\
\bar{\chi}_-(\omega)&=&\left(\frac{ND^2}{\hbar\Gamma}\right)\frac{1}{2}\frac{-i\Gamma[i(\delta-\Delta)-\Gamma_{12}]}{(i\delta-\Gamma_{e1})[i(\delta-\Delta)-\Gamma_{12}]+|G|^2},
\label{solve}
\end{eqnarray}
\end{mathletters}
where $2G=2\vec{d}_{e2}.\vec{\cal E}_c/\hbar$ is the Rabi frequency for the pump and $\Delta=\omega_c-\omega_{e2}$ is the pump detuning. 

\begin{figure}
\epsfxsize 9cm
\centerline{
\epsfbox{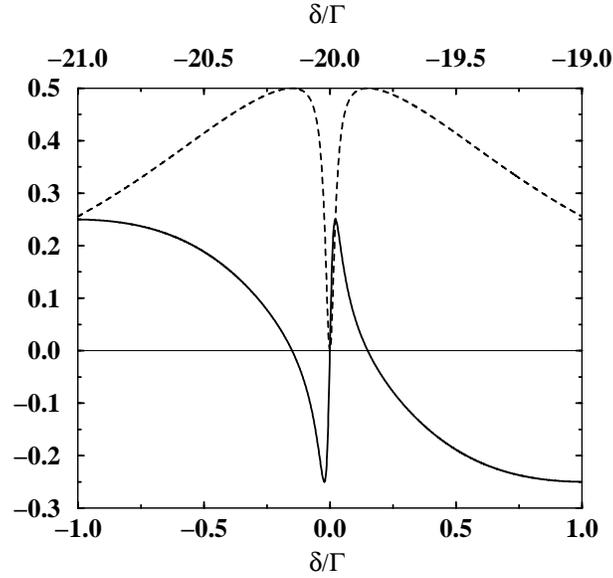}}
\caption{The variations of real (solid line) and imaginary (dotted line)  parts 
of the susceptibilities $\bar{\chi}_+$ [upper tick-labels in $x$-axis] and $\bar{\chi}_-$ [lower tick-labels in $x$-axis] in 
units of $ND^2/\hbar\Gamma$ with probe detuning $\delta/\Gamma$ are plotted 
here. The parameters used are $G=0.15\Gamma$, $B=10\Gamma$, $\Delta=0$, $\Gamma_{e1}=\Gamma_{e3}=\Gamma$, and $\Gamma_{12}=\Gamma_{23}=0$.
At the EIT window $\delta=\Delta=0$ of the $\sigma^-$ component, the 
Im$[\bar{\chi}_+]$ attains a value of $3.38 \times 10^{-7}$.}  
\label{chitune}
\end{figure}

These susceptibilities in units of $ND^2/\hbar \Gamma$ have been plotted
with respect to $\delta/\Gamma$ in the Fig.\ \ref{chitune}. We have used the parameters for $^{23}$Na vapor with $A=6.2 \times 10^7\; \textrm{s}^{-1},\;\lambda = 5890\; \rm{\AA},\;$$N=2.2 \times 10^{11}\;\textrm{atoms cm}^{-3}$. We have assumed a Zeeman 
splitting of $10\Gamma$ for the present case, which corresponds to a magnetic field of amplitude $\sim 70$ G.

\begin{figure}
\epsfxsize 9cm
\centerline{
\epsfbox{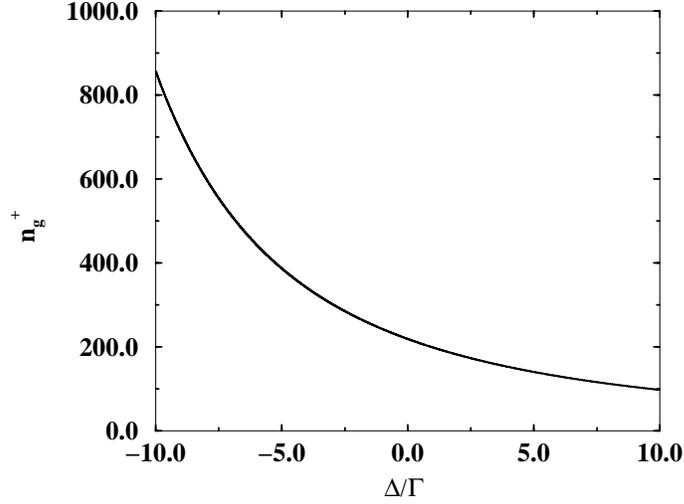}}
\caption{This figure shows the variation of $n_g^+$ with the pump detuning 
$\Delta/\Gamma$ at the EIT window $\delta=\Delta$ of $\sigma^-$ component. 
The parameters used here are $N=2.2 \times 10^{11}$ atoms cm$^{-3}$, $\lambda=
5890$ \AA, and $\Gamma= 3.1 \times 10^7$ s$^{-1}$. All the other parameters are 
the same as in the Fig.\ \ref{chitune}. Here $n_g^-$ remains constant at a value  $\sim 3.92 \times 10^6$.}
\label{detune}
\end{figure}

The system has two EIT windows at
the frequencies $\delta=\Delta$ and $\delta=\Delta - 2B$. 
The lower tick-labels in the $x$-axis in the Fig.\ \ref{chitune} shows that when $\delta=\Delta$, i.e., when
the central frequency of the pulse is near-resonance with the $|e\rangle\leftrightarrow|1\rangle$ transition,
the $\sigma^-$ component shows a normal dispersive nature, which corresponds
to a slow group velocity $v_g^-$;  whereas
the dispersion of the $\sigma^+$ component shows a flatter
behavior in frequency domain, which means that this component will propagate
 with a group velocity not too different from the velocity in vacuum. Note that because the $\sigma^+$
component is far detuned, its absorption through the medium is small ($\textrm{Im}[\bar{\chi}_+(\delta = \Delta)] \sim 3.38 \times 10^{-7}$).
The medium will appear transparent to the $\sigma^-$ component also. 
A similar situation prevails when $\delta=\Delta-2B$, i.e., when the input pulse
has a central frequency which is near-resonance 
with the $|e\rangle\leftrightarrow |3\rangle$ transition [upper tick-labels in $x$-axis; Fig.\ \ref{chitune}]. In that case,
the $\sigma^-$ component would propagate faster than the other. However both the components still propagate with 
negligible absorptions. In either case, because of the 
difference in group velocities inside the medium, the two circularly polarized 
components
 will come out of the medium at different times, without being absorbed significantly.
 Thus, the medium 
separates the two polarization components of the input pulse temporally.  
We have shown the response of the medium for the off-resonant control field in the Fig.\ \ref{detune} under the EIT condition $\delta=\Delta$. For this condition $n_g^-$ attains a value of $3.92 \times 10^6$, whereas $n_g^+$ depends on the value of $\Delta$.

\begin{figure}
\epsfxsize 9cm
\centerline{
\epsfbox{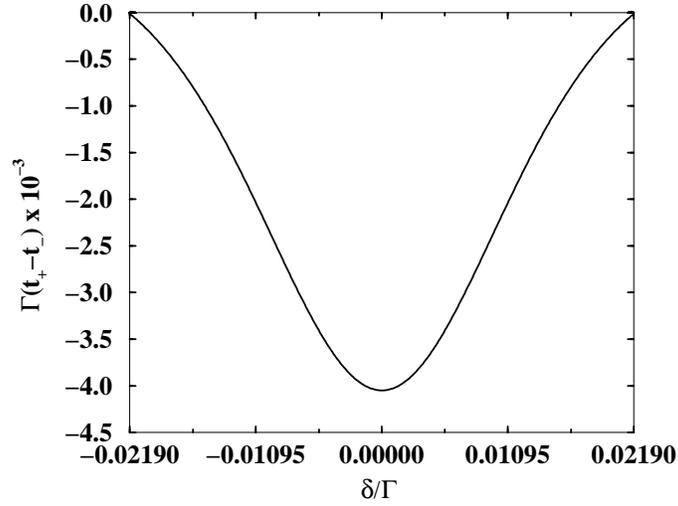}}
\caption{The variations of temporal separation between the two pulses with the 
probe detuning $\delta/\Gamma$ are shown here. 
The parameters used here are $L=1$ cm and $\Delta=0$. All the other parameters are
the same as in the Fig.\ \ref{detune}. Note that 
$\sigma^+$ component moves faster inside the medium than the $\sigma^-$ component around the EIT window $\delta=\Delta$.}
\label{timediff}
\end{figure}

For a medium of length $L=1$ cm, we have 
plotted the variation of temporal separation $\Gamma (t_+-t_-)$  
between the two polaization-components of the pulse with 
the probe-detuning $\delta/\Gamma$ in the Fig.\ \ref{timediff}. Here $t_\pm = L/v_g^\pm$ are the times taken by the $\sigma^\pm$ components to travel through the
medium. 
The maximum time separation between two components is about $-130$ $\mu$s. Clearly we can reverse the role of $\sigma^+$ and $\sigma^-$ by working at 
$\delta=\Delta-2B$.

\begin{figure}
\epsfxsize 10cm
\centerline{\begin{tabular}{cc}
\psfig{figure=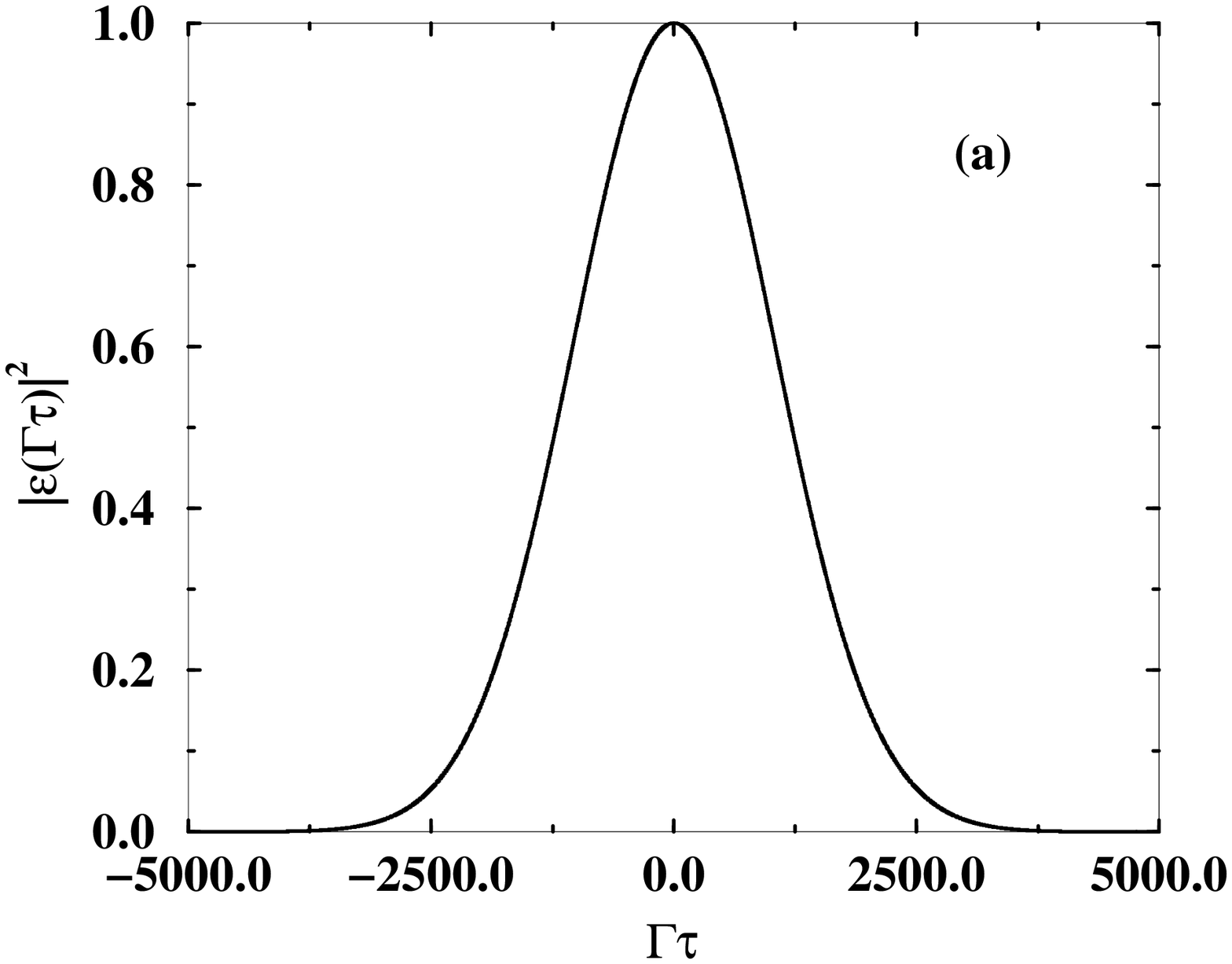,height=7cm}&
\psfig{figure=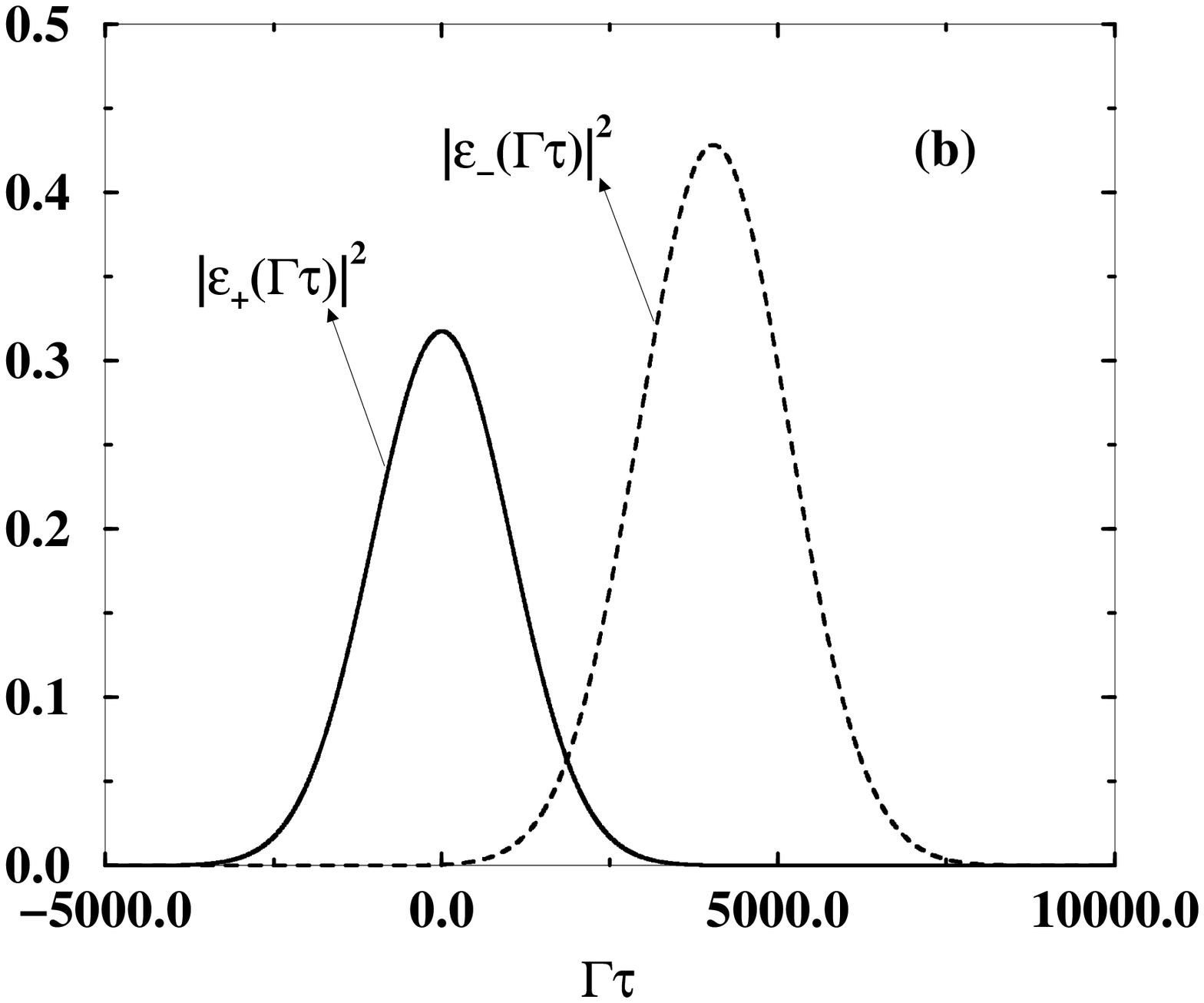,height=7cm}
\end{tabular}}
\caption{(a) This displays the input Gaussian pulse in time-domain with a width 
of $2\pi \times 4.775$ kHz;~$|{\cal E}_+|^2=|{\cal E}_-|^2=|{\cal E}|^2/2$.
(b) The two orthogonal components of the linearly polarized input probe 
pulse at the output of the medium are displayed here. It
also demonstrates the temporal separation between them for $\delta=\Delta$. The solid line shows the $\sigma^+$ component and the dashed curve refers to the 
$\sigma^-$ component. The parameters used here are the same as in the Fig.\ \ref{detune} and $\tau=t-L/c$.}
\label{pulse}
\end{figure}

We next confirm these results by studying the propagation of a Gaussian pulse 
[Fig.\ \ref{pulse}(a)] with an envelop given by 
\begin{equation}
{\cal E}(\omega)={\cal E}_0 \frac{1}{\sigma\sqrt{\pi}} \exp{[-\omega^2/\sigma^2]};~~~~~{\cal E}(t)={\cal E}_0 \exp{(-\sigma^2 t^2/4)}.
\end{equation}
For our numerical calculation, we choose $\sigma = 2\pi \times 4.775$ kHz (cf. $\Gamma = 3.1 \times 10^7$ s$^{-1}$). Using Eqs. (\ref{out}), (11a) and (11b), we evaluate numerically the output pulse and show
the results in the Fig.\ \ref{pulse}(b). 

For the chosen density and the central frequency of the input pulse, the $\sigma^+$ component does suffer absorption and
broadening. We calculated the envelop of the $\sigma^-$ component at the output 
as
\begin{eqnarray}
{\cal E}_-(L,t)&=&{\cal E}_0\frac{\sigma'}{\sigma}\exp{\left[-\frac{\sigma'^2}{4}\left(t-\frac{L}{v_g^-}\right)^2\right]};\\
\sigma'=\frac{\sigma}{\sqrt{1-i\kappa}},&&~~\kappa=\frac{\sigma^2L}{2c}\left[\frac{d^2}{d\omega^2}\left\{\omega \left[1+2\pi\chi_-(\omega)\right]\right\}\right]_{\omega=\omega_0}\nonumber.
\end{eqnarray}
The intensity of the $\sigma^-$ component thus gets reduced by Im$(\kappa)$, which depends on the second derivative of the susceptibility. The reduction as seen 
in the Fig.\ \ref{pulse}(b) is in conformity with it as the parameter Im$(\kappa)$ is about $0.15$ for the chosen parameters.

However the two components are {\it well separated in time\/}. The time separation between the two peaks in Fig.\ \ref{pulse}(b) is of the order of $4000$ in units of $1/\Gamma$, which is in agreement with the value given in the Fig.\ \ref{timediff} which is based on the calculation of group velocities. The time separation can in principle be made larger if we increase the density of the medium. However, the latter option would make the absorption of $\sigma^+$ quite large leading essentially to an output pulse which is primarily  $\sigma^-$ polarized.

Similar results can be obtained for propagation in other systems. For example, for $^7$Li \cite{zeil} in which the
 Lande' g-factor for hyperfine levels is positive, the result for pulse separation can be obtained 
 by changing $B$ to $-B$. 
The temporal separation between the
two components at the output of such a medium is about $+282.5$ $\mu$s.
This means that the $\sigma^-$ component will come out earlier for $\delta =\Delta$.

In conclusion, we have shown how a coherent anisotropic medium can be made to work  like a polarization
splitter of pulses. This is achieved by applying a coherent pump field to create
EIT for one polarization component whereas the other component suffers little absorption as it is detuned from resonance. 
 The group velocity of one of the components 
is much less than $c$,
while the other component travels through the medium without being much
affected.  Thus the orthogonal polarized components of the pulse
 get temporally separated out after passing through the medium. 

GSA thanks E. Arimondo for discussions on this paper.

\end{document}